# An analysis of the emergence of adaptive Bayesian priors from Hebbian learning in a simple attractor network model

Timothy Verstynen and Philip N. Sabes

*Department of Physiology and the Keck Center for Integrative Neuroscience*
*University of California, San Francisco 94143*

We have recently shown that the statistical properties of goal directed reaching in human subjects depends on recent experience in a way that is consistent with the presence of adaptive Bayesian priors (Verstynen and Sabes, 2011). We also showed that when Hebbian (associative) learning is added to a simple line-attractor network model, the network provides both a good account of the experimental data and a good approximation to a normative Bayesian estimator. This latter conclusion was based entirely on empirical simulations of the network model. Here we study the effects of Hebbian learning on the line-attractor model using a combination of analytic and computational approaches. Specifically, we find an approximate solution to the network steady-state. We show numerically that the solution approximates Bayesian estimation. We next show that the solution contains two opposing terms: one that depends on the distribution of recent network activity and one that depends on the current network inputs. These results provide additional intuition for why Hebbian learning mimics adaptive Bayesian estimation in this context.

ArXiv [cond-mat.dis-nn]

We have recently shown how Hebbian learning, in the context of a line-attractor neural network model, appears to approximate adaptive Bayesian estimation (Verstynen and Sabes, 2011). The conclusions in this paper were based on network simulations. Here we provide an analysis of how associative learning affects the network dynamics and why these changes mimic Bayesian estimation. The present paper serves as a supplemental note to Verstynen and Sabes (2011).

Ideally, we would want to "solve" the learning dynamics equations, yielding an expression for the network output given a sequence of inputs, but this problem is intractable (see below). Instead, we will break the problem into two parts, first quantifying the changes in the lateral connection weights $W$ due to a simplified learning scheme and then asking how these weight changes affect the steady-state activity of the network.

*Continuous dynamics and the steady-state solution.* We first consider a version of the attractor network that is continuous in both space, $c$, and time, $t$ (Wu and Amari, 2005; Wu et al., 2003):

$$\frac{d}{dt}U(c,t) = -\frac{1}{\tau}U(c,t) + \int_{c'=-\infty}^{\infty} W(c,c')X(c',t)dc' + I_c$$

$$X(c,t) = \frac{U^2(c,t)}{b\int_{c'=-\infty}^{\infty} U^2(c',t)dc'} \qquad (1)$$

These are the limits of the discrete dynamics used in Verstynen and Sabes (2011; Equations 10-11) with infinitely small iteration time-steps and infinitely many neurons with uniformly distributed preferred directions in a linear space, $c \in \mathbb{R}$. As in the discrete case, the initial value of the connection weights is assumed to be a function of the distance between the respective input and output locations,

$$W_0(c-c') = \exp\left(-\frac{(c-c')^2}{2w^2}\right). \qquad (2)$$

Wu and Amari (2005) noted that with these weights, the network has a 1-dimensional manifold of steady-state solutions,

$$U_{\infty,z}(c) = \frac{1}{b\sqrt{2}}\exp\left(\frac{(c-z)^2}{4w^2}\right), \quad X_{\infty,z}(c) = \frac{1}{b\sqrt{2\pi w^2}}\exp\left(\frac{(c-z)^2}{2w^2}\right). \qquad (3)$$

This "line attractor" is indexed by the variable z, which can be thought of as the location encoded by the activity pattern.

While the full dynamics of this network have not been solved, several papers have argued, with either empirical simulations or linearization of the dynamics, that when the input is transient, the network converges to the location parameter z that permits the greatest overlap between the input and the steady state activity patterns (Latham et al. 2003; Wu and Amari 2003). Here we make the additional observation that if the input activity is persistent and proportional to the steady-state activation pattern (Equation 3) for some location parameter $z_I$,

$$I_{z_I}(c) = A_I \exp\left(\frac{(c-z_I)^2}{4w^2}\right), \qquad (4)$$

then the network has only a single steady-state solution, the point on the manifold of Equation 3 corresponding to $z=z_I$.

*Hebbian Learning.* We now ask how learning affects the network weights $W$. We first consider the weight changes that result from experiencing just a single training trial. In order to make the problem tractable, we make several simplifying assumptions. Following Wu and Amari (2005), we assume that the network relaxation time is very fast compared to the trial duration, so that we only need to consider the steady-state values of the network in determining the activity. Second, we assume that the input is noise-free and takes the form of Equation 4, so that Equation 3 fully describes the steady-state. Next, we assume that the intra-trial effects of learning are negligible, so that learning can be considered as a discrete, per-trial process. Finally, since we are only considering a single trial of learning, we assume a simple, un-normalized Hebbian rule. With these assumptions, the change in weights on a single trial is given by

$$\Delta W_{z_0}(c,c') = \beta X_{\infty,z_0}(c) X_{\infty,z_0}(c')$$
$$= \frac{\beta}{b^2 2\pi w} \exp\left(-\frac{(c-c')^2}{2w^2} - \frac{(c-z_0)(c'-z_0)}{w^2}\right) \quad (5)$$

where $z_0$ is the location parameter of the input on that trial. Note that the weight changes are both local to the diagonal of the weight matrix (first term), i.e. connections between neighboring units are primarily affected, and local to the trained location (second term), i.e. units with preferred directions near $z_0$ are primarily affected.

Next, we consider the effect of several training trials where the input parameter $z_I$ is selected from a normal distribution $N(z_0, \sigma^2_{targ})$. Here $\sigma^2_{targ}$ is the correlate of the context variance in our experiments. Because the influence of learning is cumulative across trials, there is no closed-form expression for the weight changes $\Delta W$ that would result from a sequence of trials drawn from this distribution. Instead we solve for the expected value of the weight change from a single trial, given this input distribution:

$$\Delta W_{z_0,\sigma^2_{targ}}(c,c') = E\left[\Delta W_z(c,c') \mid z_0, \sigma^2_{targ}\right]$$
$$= \frac{\beta}{b^2 2\pi w}\left(\frac{w^2}{w^2+\sigma^2_{targ}}\right)^{\frac{1}{2}} \exp\left(-\frac{1}{w^2+2\sigma^2_{targ}}\left((c-c')^2 \frac{w^2+\sigma^2_{targ}}{2w^2} + (c-z_0)(c'-z_0)\right)\right) \quad (6)$$

Equation 6 reduces to the single-trial case of Equation 5 when $\sigma^2_{targ} = 0$. As $\sigma^2_{targ}$ increases, the weight change gets smaller and more diffuse. Equation 6 can also be thought of as the effect of a series of trials, under the assumption that the weight changes have no effect during that sequence, i.e. under a batch learning model.

*Post-learning steady-state.* As stated above, there is no closed-form solution for the steady-state activity of the network given the weight changes defined above. While further analysis of the steady-state is provided later, we first use numerical simulations to quantify the effects of these weight changes. We simulated the discrete version of the network (Equations 10-11 in Verstynen and Sabes, 2011) with weight matrix $W = W_0 + \Delta W_{z_0,\sigma^2_{targ}}$ and a noise-free input with gain $A_I$ and location parameter $z_I$ (see Equation 4). This network rapidly converges to a steady-state pattern of activity. We computed the effective location parameter of this activity, $\hat{z}$, using population vector decoding, as above (Equation 13, in Vestynen and Sabes, 2011) and the relative bias of this read out, defined as $(z_I - \hat{z})/(z_I - z_0)$. The solid lines in Figure 1 show the relative bias as a function of the input gain for a variety of target distributions $\sigma_{t\arg}$ and for a small separation between training and input locations ($z_I - z_0 = w/16$). When $A_I = 0$, the input has no effect, so $\hat{z} = z_0$ and the relative bias is equal to unity. As $A_I$ gets larger, the input has a greater effect and the relative bias falls. For a given input gain, the relative bias decreases with increasing $\sigma_{t\arg}$, reflecting the decreased effects of the smaller and more diffuse weight changes (see Equation 6).

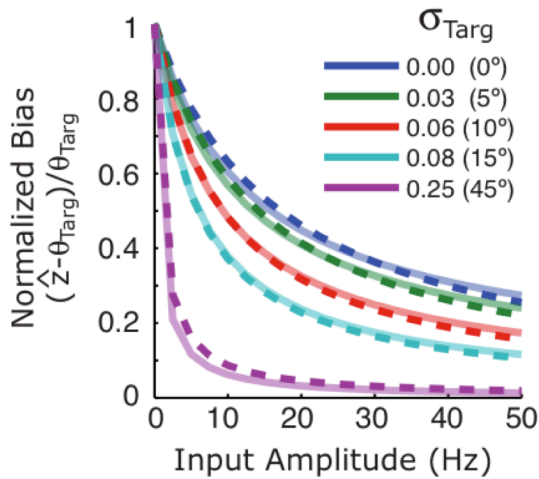

**Figure 1.** Normalized bias of the simplified, noise-free network (solid lines) with different input gains on the probe trial and following learning with different context variances, $\sigma^2_{t\arg}$. Dashed lines represent predictions from the simple Bayesian model after fitting to the network output with two free parameters.

We next compared the pattern of biases observed in our simulations to those expected from a simple Bayesian (MAP) estimator with normal prior and likelihood distributions. Motivated by results from the full network simulation (Figure 8 in Verstynen and Sabes, 2011) and previous reports (Latham et al. 2003; Ma et al. 2006), we assumed that the likelihood variance was proportional to the inverse of $A_I$. The only remaining free parameters are the mapping from context variance $\sigma^2_{t\arg}$ to the variance of the prior distribution. We assumed a linear relationship, and fit the two free parameters to the simulation results (solid lines in Figure 1). The resulting bias predictions are shown in the dashed lines of Figure 1. The quality of the fit is striking, given that only two free parameters were used. This correspondence illustrates that even in this very simplified Hebbian learning paradigm, the bias effects are consistent with the presence of a Bayesian prior.

*Analysis of the steady-state.* We can't solve the dynamics of Equations 1 with the new weights of Equation 5. However, we can get a strong intuition of how these dynamics give rise to a Bayesian-like process by making a few additional approximations. First, we approximate the steady state solution using the same functional form as the pre-learning steady-state, but with unknown width parameter $w_s$, location parameter $z_s$, and activation gain $A_s$,

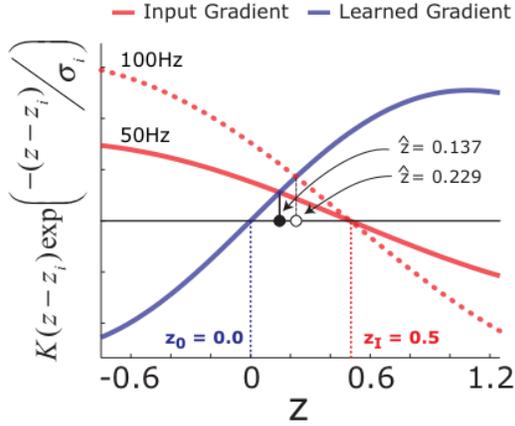

**Figure 2.** Illustration of the solution to the approximate steady-state equation (Equations 8-9). The blue curve corresponds to the effects of learning on the previous trial, trained at input parameter $z_0$. The solid red curve shows the effects of the current input at $z_I$ for an example input gain. Their intersection (●) is the approximate steady state solution. The dashed red line shows the effect of doubling the input gain, which shifts the solution (○) closer toward $z_I$.

$$U_s(c) = A_s \exp\left(\frac{(c-z_s)^2}{4w_s^2}\right), \quad X_s(c) = \frac{1}{b\sqrt{2\pi w_s^2}} \exp\left(\frac{(c-z_s)^2}{2w_s^2}\right). \tag{7}$$

Next, instead of solving for the true-steady state, we instead look for a value of $z_s$ for which the time derivative of the activation function $dU_s/dt$ is orthogonal to the line attractor, i.e. is orthogonal to the vector $dU_s/dz_s$:

$$0 = \int_{c=-\infty}^{\infty} \frac{dU_s(c)}{dz_s} \frac{dU_s(c)}{dt} dc, \tag{8}$$

where $dU_s/dt$ comes from Equation 1. With some algebraic manipulation, this equation simplifies to the form

$$0 = K_0(z-z_0)\exp\left(\frac{-(z-z_0)^2}{s_0}\right) - K_I(z_I-z)\exp\left(\frac{-(z-z_I)^2}{s_I}\right). \tag{8}$$

When the new weights, $W = W_0 + \Delta W_{z_0, \sigma_{targ}^2}$, are used, the parameters in Equation 8 are

$$K_0 = \frac{\beta}{b^3 \pi^{1/2}} \frac{w_s(w^2+w_s^2)}{\left(\frac{1}{2}S_o(2w^2+3w_s^2)\right)^{3/2}} \qquad K_I = A_I \frac{4\pi^{1/2}}{S_I(w_1^{-2}+w_I^{-2})^{1/2}}$$

$$S_0 = 2\left(\sigma_{targ}^2 + \frac{1}{(w^2+2w_s^2)^{-1}+(w^2+w_s^2)^{-1}}\right) \qquad S_I = 4(w_1^2+w_I^2) \tag{9}$$

The solution to Equations 8-9 is illustrated for a sample case in Figure 2. Note that the equation is a difference of two terms, each taking the form of the derivative of a Gaussian, centered at $z_0$ and one at $z_I$, respectively. These terms only have the same sign in the interval between $z_0$ and $z_I$, and so the solution must lie in that interval. The exact location where the two terms are equal (circles in Figure 2) is determined by the parameters in S9. There are three key variables to consider. First, the input parameter $A_I$ determines the gain of the second term. When $A_I$ increases, so does $K_I$, the gain of the red curve (dashed red line), driving the intersection the two curves rightward, i.e. driving the solution (open circle) closer to $z_I$. Similarly, either an increase in the learning rate β or a decrease in the context variance $\sigma^2_{t\arg}$ would lead to an increase in $K_0$, the gain of the blue curve, pushing the solution towards $z_0$. These examples illustrate the competition between the effect of learning (blue curve) and the effect of the current input (red curve), and how the input gain, learning rate, and context variance effect this competition. It is this interaction between these parameters that drive the Bayesian-like behavior of the network.

Note however that the solutions identified above are only an approximation. The reason for this is that the true steady-state activity patterns $U(c)$ and $X(c)$ do not take the form Equation 8. In fact, there is no simple closed-form expression for these patterns, and empirical tests show that the solution to Equation 8 underestimates the bias of the steady-state readouts. More importantly, however, the effects of $A_I$, β, and $\sigma^2_{t\arg}$ predicted by this simple analysis are borne out in the numerical simulations (e.g., Figure 1), and capture the same trends observed in the full network model (Figure 8 in Verstynen and Sabes, 2011). As in the full network model, we see in the simplified network a critical role for noise during the readout (post-learning) trials. In fact, if we add input noise to the simplified simulations that generated Figure 1, the bias curves look much more like those of the full simulations with higher baseline rates (i.e, Figure 8 in Verstynen and Sabes, 2011; comparison data not shown).

Finally, we contrast these results with those of Wu and Amari (2005), who also argue that associative learning can mimic Bayesian estimation in this same attractor network, but with a different learning rule. In that paper, the authors approximated a general solution to the learning dynamics using two key approximations: i) the inputs are small with negligible noise and ii) the effects of cumulative learning are small, so that the original steady-state solution can be used to model the effects of later trials. The assumption of small inputs is not appropriate for our study, since we left the inputs on throughout the simulated trial. We also make the assumption that the cumulative effects of learning are small, however we make that more explicit by only attempting to model the effects (or expected effects) of a single trial of learning. Nonetheless, our goal was somewhat different from theirs. Rather than trying to show explicitly that network mimics Bayesian inference (which is intractable for our model), we wanted a simplified framework to study in more detail how the parameters of the inputs ($\sigma^2_{t\arg}$, $A_I$) and the learning (β) effectively balance the tradeoff between the effects of learning and the effect of the next input.

**Acknowledgements.** This work was supported by NIH Grant P50 MH077970 (Conte Center) and the Swartz Foundation.